# Quantum confining excitons with electrostatic moiré superlattice

Liuxin Gu[1], Lifu Zhang[1], Sam Felsenfeld[2], Rundong Ma[3], Suji Park[4], Houk Jang[4], Takashi Taniguchi[5], Kenji Watanabe[6], You Zhou[1,7,†]

[1]Department of Materials Science and Engineering, University of Maryland, College Park, MD 20742, USA
[2]Department of Physics, University of Maryland, College Park, MD 20742, USA
[3]Department of Electrical and Computer Engineering, University of Maryland, College Park, MD 20742, USA
[4]Center for Functional Nanomaterials, Brookhaven National Laboratory, Upton, NY 11973, USA
[5]Research Center for Materials Nanoarchitectonics, National Institute for Materials Science, 1-1 Namiki, Tsukuba 305-0044, Japan
[6]Research Center for Electronic and Optical Materials, National Institute for Materials Science, 1-1 Namiki, Tsukuba 305-0044, Japan
[7]Maryland Quantum Materials Center, College Park, Maryland 20742, USA

†To whom correspondence should be addressed: youzhou@umd.edu

## Abstract

Quantum confining excitons has been a persistent challenge in the pursuit of strong exciton interactions and quantum light generation. Unlike electrons, which can be readily controlled via electric fields, imposing strong nanoscale potentials on excitons to enable quantum confinement has proven challenging. In this study, we utilize piezoresponse force microscopy to image the domain structures of twisted hexagonal boron nitride (hBN), revealing evidence of strong in-plane electric fields at the domain boundaries. By placing a monolayer $MoSe_2$ only one to two nanometers away from the twisted hBN interface, we observe energy splitting of neutral excitons and Fermi polarons by several millielectronvolts at the moiré domain boundaries. By directly correlating local structural and optical properties, we attribute such observations to excitons confined in a nanoscale one-dimensional electrostatic potential created by the strong in-plane electric fields at the moiré domain boundaries. Intriguingly, this 1D quantum confinement results in pronounced polarization anisotropy in the excitons' reflection and emission, persistent to temperatures as high as ~80 Kelvins. These findings open new avenues for exploring and controlling strongly interacting excitons for classical and quantum optoelectronics.

**Introduction**

Confining and manipulating excitons at the nanoscale is a long-standing goal of both fundamental and technological importance[1–7]. Quantum confinement of excitons enables fundamental studies of strongly interacting quantum many-body systems, such as strongly interaction excitons and quantum light generation[1–7]. Additionally, confining and manipulating exciton flow may lead to novel devices for energy transfer and information processing[8,9]. However, unlike electrons, which can be readily controlled via an electric field[10], imposing strong nanoscale potential on excitons for quantum confinement has been challenging[9,11].

Substantial progress has been made in confining excitons in van der Waals heterostructures based on atomically thin semiconductors[4,5,7,12]. Notably, quantum confinement of excitons has been achieved in the moiré patterns of two-dimensional semiconductors, with a periodicity varying from hundreds to a few nanometers[13–18]. This moiré potential, driven by interlayer electronic hybridization, can be remarkably strong, splitting the exciton energies by tens of meVs through quantum confinement[18,19]. However, interlayer hybridization induces lattice relaxation and often a transition to an indirect band gap, which causes undesirable effects such as exciton linewidth broadening and reduced quantum yield[20–23]. Alternatively, a recent approach uses the fringe in-plane electrical field of patterned gate electrodes to directly modulate the exciton energies without changing the materials' band structures[5,7,12]. Compared with moiré potential, the strength of the confinement is weaker, primarily limited by the gating geometry and electrical breakdown of dielectrics.

Here, we develop a method that takes advantage of both methods to realize strong confinement of excitons while maintaining the direct band gap. We leverage the recently discovered moiré ferroelectricity[24–27] in twisted hexagonal boron nitride (hBN) to imprint a nanoscale potential on excitons in 2D semiconductors[27–31]. Different from these earlier studies focusing on the effects of out-of-plane electric fields inside moiré domains[32,33], we utilize the strong local in-plane electric fields at the moiré domain boundary to confine excitons. By placing the 2D semiconductor just a few atoms away the twisted hBN interface, we confine excitons inside a narrow strip with a width of ~ten nanometers, which leads to a highly polarized exciton response. These results open exciting avenues for studying many-body quantum systems, such as strongly interacting excitons and photons[34], and novel ferroelectric control mechanisms for classical and quantum optoelectronics[35–37].

**Results**

To create moiré superlattice domains, we stack two thin layers of hBN with a near-0° target twist angle. The local variations of twist angles create domains of varying sizes, which we use to correlate structural and optical properties[27,36]. In such a twisted hBN structure, triangular AB and BA stacking domains, with broken inversion symmetry, generate out-of-plane polarizations in opposite directions (Fig. 1a). Near the domain boundaries, Gauss's law dictates an in-plane electric field must be present. The magnitude of such an in-plane field within the interface, arising from the slight atomic-scale sliding of the two layers, is predicted to be quite strong, comparable to the out-of-plane polarization in the AB/BA domains[38,39]. As the vertical distance $d$ to the twisted interface increases, we expect a decrease in the in-plane electric field[27].

To create a strong electric field, we select hBN layers thinner than ~2 nm so the twisted interface remains close to the sample surface and characterize the surface with piezoresponse force microscopy (PFM). Unlike Kelvin probe force microscopy, which is sensitive to variations in surface potential, such as those arising from out-of-plane polarizations[24,27,32,40–42], PFM detects sample deformation through the converse piezoelectric effect. In this approach, both out-of-plane and in-plane polarizations can be detected from the different modes of cantilever movements, such as vertical displacement, buckling, and torsion[43,44], distinguished by their respective photodiode responses[45] (Figs. 1b,c).

Here, we focus on the in-plane polarization at the twisted hBN domain boundaries. Figures 1d, e present the measured PFM amplitude and phase maps, revealing the clear formation of multiple domains and domain boundaries. Domain sizes vary significantly across different regions, ranging from a few tens of nanometers to several micrometers, due to local variations in the twist angle, despite the sample being fabricated at a nominally zero twist angle. Interestingly, we observe the strongest piezoelectric response at the domain boundaries with weaker contrast across the two types of domains, in stark contrast to KPFM studies of similar systems[44]. Furthermore, the PFM response at the domain boundary is most pronounced when the boundary aligns parallel to the cantilever and can be positive or negative depending on boundary orientations.

These observations reveal a strong in-plane electric field at the domain boundaries. As shown in Figs.1b, c, when the domain boundary is parallel to the cantilever, the in-plane polarization perpendicular to the boundary induces torsion of the cantilever, generating a PFM signal. The sign of the PFM signal at the domain boundary therefore reflects the different directions of the in-plane electric field. When the in-plane field is parallel to the cantilever, the cantilever buckles, which is detectable via vertical motion and has been recently reported[43,46]. In our experiments, however, this corresponding vertical signal is not selected by the photodiode, further evidenced by the uniform amplitude across different domains.

Next, we integrate a monolayer $MoSe_2$ flake on top of the twisted hBN and encapsulate it with additional top hBN and graphene. The entire fabrication process was performed without heating the samples to a high temperature to avoid perturbation of the domain structures. We then perform the reflectance and PL measurements of the sample at 5 K to establish a correlation between the twisted hBN domain sizes and the $MoSe_2$ exciton spectroscopic features.

We first characterize excitons in regions with large domains of ~1-2 μm shown in Fig.2a, where a single domain can be resolved in the far field. Figures 2b, c show the corresponding reflectance spectrum of the $MoSe_2$ taken within the region. When the laser is within a single domain, both PL and reflection spectra show a single exciton peak, similar to $MoSe_2$ encapsulated in non-twisted hBN[47]. However, when the laser moves on to the domain boundaries, we observe a split in the exciton energy of around ~3 meV in both PL and reflectance, dramatically different from typical hBN-encapsulated monolayers.

We further investigate how the energy splitting varies as a function of domain size across the sample. As shown in Fig. 2 d, e, we observe that exciton energy splitting increases as domain size decreases, reaching as high as ~7 meV when the domain size is approximately 250 nm. Since these domain sizes are below the diffraction limit, the measured exciton response includes contributions from both within the domains and along the domain walls. Interestingly, the exciton splitting also correlates with the strength of the in-plane electric field. By comparing regions with similar domain sizes but different in-plane electric fields, as indicated by PFM amplitude, we observe larger exciton splitting in areas with stronger in-plane fields (see difference in Fig. 3d and Fig. S2).

Several mechanisms can influence the excitonic spectra of MoSe$_2$ on twisted hBN. For instance, out-of-plane polarization can introduce varying doping levels across different domains[31,48–50]. While this may shift the exciton energy, a pure doping effect would not cause exciton splitting, since the doping—and thus the exciton resonance—are expected to change continuously across domains[31]. In contrast, inhomogeneous strain can shift and split exciton energy[53]. This effect can occur even in TMDs encapsulated in non-twisted hBN, often arising uncontrollably in areas with strong strain gradients, such as near bubbles[7,54].

Another possible mechanism is related to the in-plane electric field $\boldsymbol{F_x}$ at the domain boundaries. In particular, modifies the exciton energy by $\Delta E = -\frac{1}{2}\alpha |F_x|^2$, where $\alpha$ is the polarization coefficient of excitons[7,52]. Therefore, a strong and spatially localized $\boldsymbol{F_x}$ at the moiré domain boundaries can create a nanoscale confinement potential $V_x$ for excitons, which induces exciton energy splitting[7,55](Fig. 2f). Observing exciton splitting exclusively near domain boundaries in large-domain regions suggests that this electrostatic effect may play a significant role in our studies. This also explains the larger exciton splitting observed at regions with stronger PFM response and smaller domains, where smaller domains likely correspond to narrower domain walls and tighter confinement.

Excitons confined in 1D potential, such as along moiré domain walls, are expected to exhibit strong anisotropic optical properties[56] (Fig. 3a). Therefore, to test our hypothesis further, we measure how reflectance and PL spectra of the excitons vary as a function of the linear polarization angle. First, we study the region with large domains so that only a single domain boundary contributes to the optical signal in the far-field (Fig. 3b). As shown in Figs. 3c, d, we observe anisotropic exciton reflection and emission with high degrees of linear polarization. By mapping the polarizer direction to the real-space angles on the sample, we find that the exciton polarization directions are parallel to the domain boundaries and remain so as we move onto domain boundaries of different spatial orientations across the sample (Figs. 3 & S2). In contrast, we did not observe any polarization-dependent behaviors in regions with a single exciton resonance without splitting (Fig. S3e). Similar exciton splitting and linear polarization behaviors are also observed in more than four different samples with similar structures (see Fig. 4a for device D2 as an example).

The observed polarization dependence can be attributed to the 1D confinement at the domain boundary. In monolayer TMDs, due to electron-hole exchange interactions, the exciton dispersion splits into longitudinal and transverse branches, which have a coherent superposition of K and -K valley excitons at zero momentum[57–59]. A 1D confinement potential breaks the $C_3$ in-plane rotation symmetry and splits the longitude and transverse branch at the zero momentum, which couples to light in two orthogonal directions[54] (Fig. 3a). With sufficiently strong confinement, this $x$–$y$ polarization energy splitting surpasses the confinement energy, resulting in all confined modes becoming linearly polarized, as observed in our experiments. This linear polarization in reflection and emission also strongly suggests that inhomogeneous strain is not the dominant factor here. In control experiments observing exciton splitting on non-twisted hBN substrates, the two strain-split exciton resonances have random polarization angle dependence, likely due to variations of the local strain (Fig. S4).

Further evidence can also be found in regions where the domain size falls below the diffraction limit. In these regions, the exciton reflectance and emission are no longer linearly polarized but still exhibit an anisotropic response. Notably, in specific spots, two local maxima of PL emission are observed at polarization angles separated by roughly sixty degrees (Fig. S3d). These phenomena can be understood from the non-perfect triangle shapes of domain boundaries measured from PFM, where one particular domain wall direction would contribute more to the measured optical signal than the other directions.

Lastly, we investigate the quantum confinement of charged excitons near the moiré domain boundaries in another gated MoSe$_2$ on hBN device, **D2**. As we vary the doping levels through electrostatic gating, we also observe a splitting in both the repulsive and the attractive polarons (Figs. S5a,b). Furthermore, the reflectance of the neutral exciton, repulsive polaron, and attractive polaron all exhibit linear polarization aligned with the same direction, parallel to the domain wall directions, indicating that 1D confinement can also occur for the Fermi polarons (Figs. 4a-c). Intriguingly, as doping density increases, the split between the repulsive polarons tends to decrease, whereas that between the attractive polarons slightly increases (Fig. S5c). A quantitative understanding of the confinement of Fermi polarons requires further theoretical studies, for instance, on the polarizability of attractive vs. repulsive polarons[7,54]. Nevertheless, such behaviors stand in stark contrast to strain-induced energy splitting, which is expected to be independent of doping.

To further demonstrate the robustness of this confinement, we measure the polarization-resolved reflectance of the neutral excitons at different temperatures. The anisotropic excitonic response persists but becomes weaker with increasing temperature. At a temperature around 80 K, the exciton splitting disappears, and the reflection and emission from excitons become isotropic in the 2D plane. This transition occurs at a temperature where the thermal energy is comparable to the energy splitting associated with the quantum confinement potential, which is consistent with our interpretation. We note that the disappearance of the splitting is not simply due to the linewidth change because the total linewidth does not change dramatically with increasing temperature up to 100 K (Fig.S6).

**Discussion and outlook**
Finally, we quantify the strength and spatial extent of the quantum confinement potential. The potential depth induced by the in-plane electric field is given by $\Delta E = -\frac{1}{2}\alpha|F_x|^2$, where $\alpha \sim 6.5$ eV nm$^2$/V$^2$ for intralayer excitons in TMDs[56–59]. From the exciton energy splitting, we estimate a lower bound for the in-plane electric field of ~33±1 mV/nm in regions of our device with the splitting energy of ~3.5 meV. This value is higher than those reported in recent experiments[30,41], likely due to the reduced separation between the TMD and the twisted hBN interface in our device. Additionally, we estimate the lateral size of the confinement to be approximately 13 nm from the simple harmonic oscillator approximation (see Methods). This estimated size is slightly smaller than the moiré domain wall size in the PFM images, possibly due to the limited PFM resolution.

Nevertheless, these results indicate that such electrostatic moiré potential imposes stronger and tighter confinement than in partially gated devices[60–63].

The demonstrated quantum confinement of intralayer excitons at moiré domain walls opens exciting new avenues for controlling excitons at the nanoscale. Such a strong in-plane electric field is expected to occur in many types of twisted hetero-structures with broken inversion symmetry[3,64], which can be readily integrated with all sorts of 2D semiconductors to engineer their electronic and excitonic properties[60–63]. The ferroelectric response observed in many of such twisted structures could enable dynamic electrical control of 1D exciton chains[32,41,64]. A particularly exciting direction is to explore how strong exciton confinement can enhance the system's nonlinearity[3,65], especially in the regime where the confinement size is comparable to the exciton blockade radius[66,67]. Quantum confining Rydberg excitons is therefore especially promising, thanks to their large polarization and blockade radius[61,68]. To access this intriguing regime, new optical methods may be required to investigate how light interacts with the 1D excitonic chains as it propagates along and within the domain walls, such as near-field techniques[69–73]. In addition, embedding such quantum confined excitons with large oscillator strengths into cavities can further enhance the photon-photon nonlinearity for quantum information processing[8,74–77].

**Acknowledgements**
We acknowledge valuable discussions with Felipe H. da Jornada. This research is primarily supported by the U.S. Department of Energy, Office of Science, Office of Basic Energy Sciences Early Career Research Program under Award No. DE-SC-0022885. The fabrication of samples is supported by the National Science Foundation CAREER Award under Award No. DMR-2145712. Funding for the AFM shared facility used in this research was provided by NSF under award number CHE-1626288. This research used Quantum Material Press (QPress) of the Center for Functional Nanomaterials (CFN), which is a U.S. Department of Energy Office of Science User Facility, at Brookhaven National Laboratory under Contract No. DE-SC0012704. K.W. and T.T. acknowledge support from the JSPS KAKENHI (Grant Numbers 20H00354, 21H05233 and 23H02052) and World Premier International Research Center Initiative (WPI), MEXT, Japan for hBN synthesis.

**Data availability**
Source data are provided with this paper. All other data are available from the corresponding authors upon reasonable request.

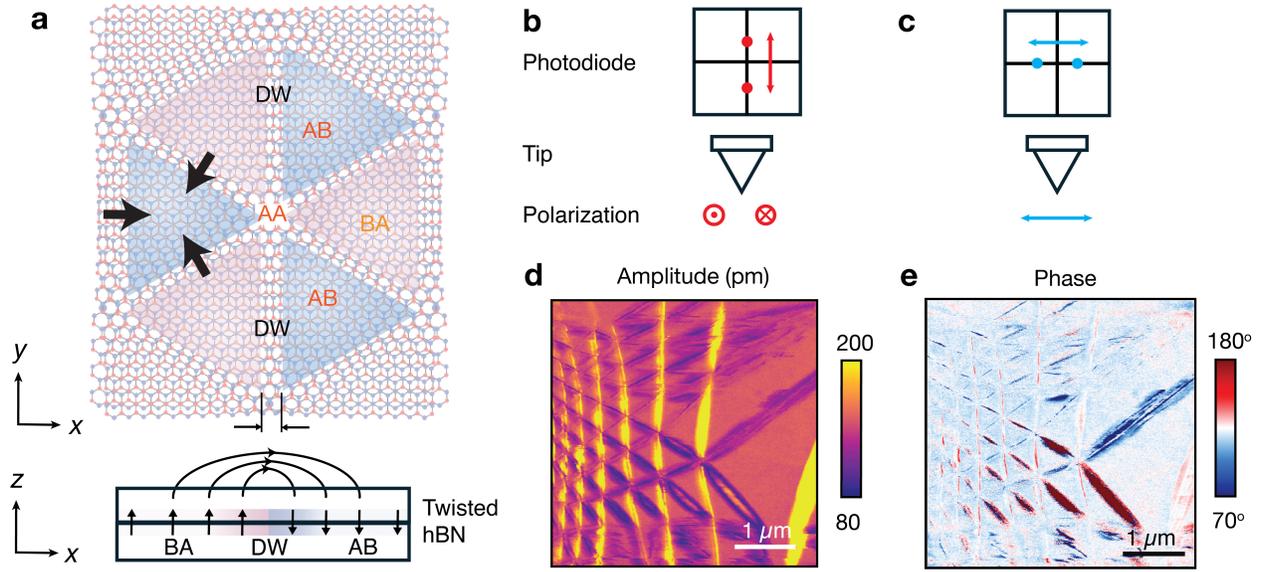

**Figure 1. In-plane electric field in twisted hBN triangular Moiré superlattice. a**, Schematic of the in-plane electric field generated at the domain boundary in twisted hBN moiré superlattice. The top panel shows the top view of the superlattice. The black arrow refers to the in-plane polarization. The bottom shows the sideview of the twisted hBN with alternating AB and BA domains with opposite out-of-plane polarization. The electrostatic potential difference across the domains creates a large in-plane electric field at the domain boundary, which decays with increasing distance from the interface. **b, c**, Different modes of the PFM cantilever displacement in response to the in-plane electric field and corresponding photodiode response (top). **b**, When the in-plane polarization is parallel to the tip axis, the tip deflects vertically. **c**, When the in-plane polarization is perpendicular to the tip axis, a lateral deflection is recorded. **d, e**, PFM amplitude and phase map of the triangular superlattice region with gradually changing domain size.

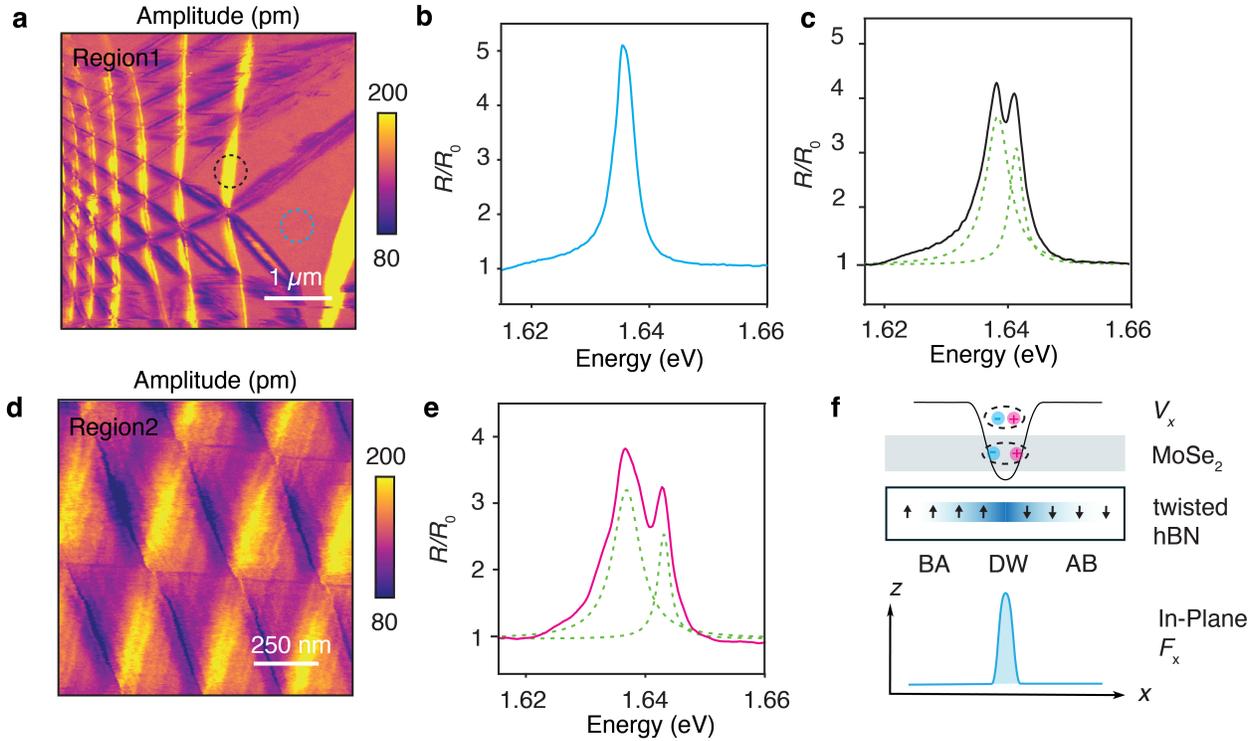

**Figure 2. Confined intralayer excitons by the in-plane electric field. a,** PFM image of the electrostatic superlattice showing different sizes of triangular domains. **b**, Normalized reflectance taken at the large domain area indicated as the blue spot in **a**. **c,** Normalized reflectance taken at the domain boundary indicated as the black spot in **a.** Green dashed lines are Lorentzian fit, and the solid line is measured data. **d**, PFM image of a region with sub-diffraction-limit domains. **e,** Normalized reflectance spectra taken at the region in **d**. **f,** Schematic of the field-induced confinement potential $V_x$ confining intralayer excitons in MoSe$_2$ adjacent to the twisted hBN at the domain boundary.

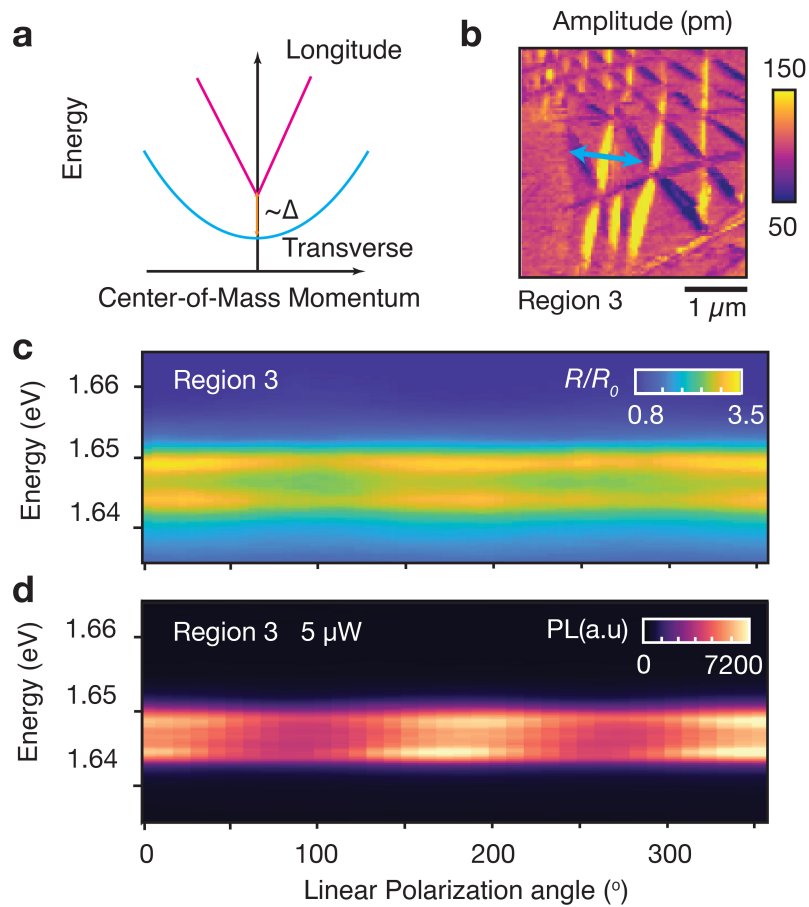

**Figure 3. 1D confinement at the domain boundary. a,** Schematic of the confined exciton dispersion. Under spatial localization, the exciton longitude and transverse branch become split with a magnitude of ∼Δ at zero center-of-mass momentum. **b,** PFM map of region 3. **c,** Normalized reflectance as a function of the excitation linear polarization angle in region 3 as indicated by the blue arrow in **b**. **d,** PL emission as a function of linear polarization with 5 $\mu W$ excitation power from 635nm CW laser. The spot is the same one taken in **c**.

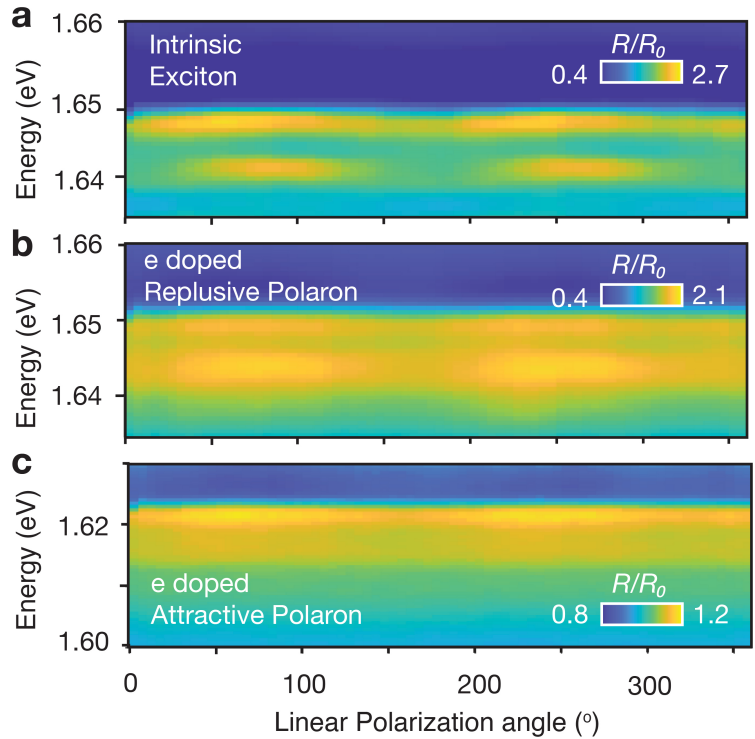

**Figure 4. Anisotropic optical response of Fermi polarons. a-c**, Linear polarization dependent reflectance of **a,** neutral exciton (gate voltage $V = -2.5$ V), **b,** repulsive polaron, and **c,** attractive polaron (gate voltage $V = 0.5$ V) at electron-doped in $MoSe_2$ in device **D2** at 5 K.

# Supplementary Information

**This file includes:**

**Materials and Methods**

**Supplementary Figures 1-7**

## Materials and Method

### Device fabrication

Graphite, hBN flakes are mechanically exfoliated from the bulk crystals onto the silicon chip with $SiO_2$ layer. Exfoliated monolayer $MoSe_2$ flakes were provided by the Quantum Material Press (QPress) facility in the Center for Functional Nanomaterials (CFN) at Brookhaven National Laboratory (BNL). The thickness of the gated hBN flakes and $MoSe_2$ layer numbers are estimated based on the color contrast under optical microscopy. Twisted hBN flake thickness is identified by contact-mode AFM. The heterostructure is assembled in a transfer station built by Everbeing Int'l Corp., which uses PDMS (Polydimethylsiloxane) and PC (Polycarbonate) as stamp and transfer all the flakes in a dry transfer method onto a silicon chip with 285nm $SiO_2$ layer. Then the electrical contacts are patterned by electron-beam lithography and a liftoff process where we deposited 5nm of Cr and 80nm of Au by thermal evaporation.

### PFM measurement

PFM characterization is performed with DART- PFM (Dual AC Resonance Tracking Piezo Force Microscopy) mode by Oxford Asylum Cypher ES at room temperature. We used Ti/Ir coated conductive cantilever probes with resonance frequency of 75 kHz and a spring constant of 2.8 N/m (ASYELEC-01-R2). We applied an AC bias with a driving amplitude of 3 V between the tip and the sample to generate an electric field in the vertical direction. This field induces a piezoelectric response in the sample, which is detected by monitoring the tip deflection through the laser diode. PFM images are obtained with a scan rate of 1Hz and a 0° scan angle.

### Optical spectroscopy

The optical measurements were performed in our home-built confocal microscope with Attodry 4K cryostat. The apochromatic objective equipped in the chamber has numerical aperture NA=0.82. The PL measurement is performed with a 635nm diode laser excitation. The reflectance measurement is performed using either a halogen lamp (from Thorlabs) or a supercontinuum white laser (from YSL Photonics Inc.) as the excitation source. The excitation and detection are both diffraction-limited, with a FWHM spot size of $\frac{\lambda}{2NA}$ close to ~500 nm. The white laser has a pulse duration of ~100 ps with a variable repetition rate of up to 40 MHz. The spectra are measured by a Horiba iHR320 spectrometer using a 600 mm/line grating and a Synapse-Plus back-illuminated deep depletion CCD camera. Polarization-dependent reflectance is measured by putting a linear polarizer right in front of the sample. For polarization-resolved PL, we add an additional linear polarizer and half-waveplate on the collection path so that the polarization onto the CCD is fixed to avoid the impact of the polarization-dependent collection efficiency of the CCD.

### Estimation of the Confinement Potential

We estimate the in-plane confinement potential based on the harmonic oscillator approximation with the exciton splitting energy. As the confinement length $l \sim \frac{\hbar \pi}{\sqrt{m_X \Delta E}}$, $m_X \sim 1.3 m_e$ and discrete energy splitting $\Delta E$ is 3.5 meV.

# Supplementary Figures

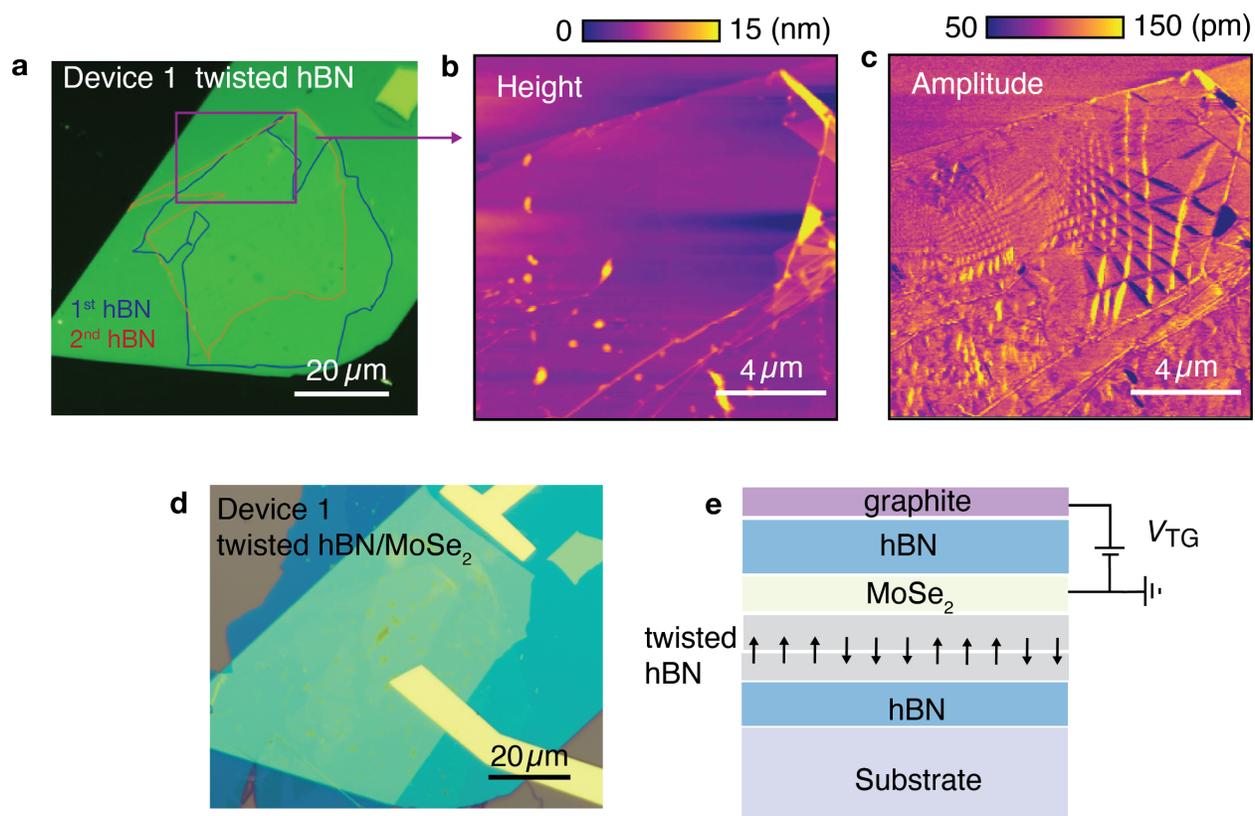

**Figure S1**. **Optical image and PFM map of Device D1**. **a**, Optical image of the R-stacked twisted hBN. **b,** Topography and **c,** PFM amplitude image of the region indicated by the purple box in **a**. After PFM characterization, we stack the monolayer MoSe$_2$ on top of the twisted hBN surface and then encapsulate it with gate hBN and top graphite. **d,** The optical image, and **e,** the device schematic of the final device.

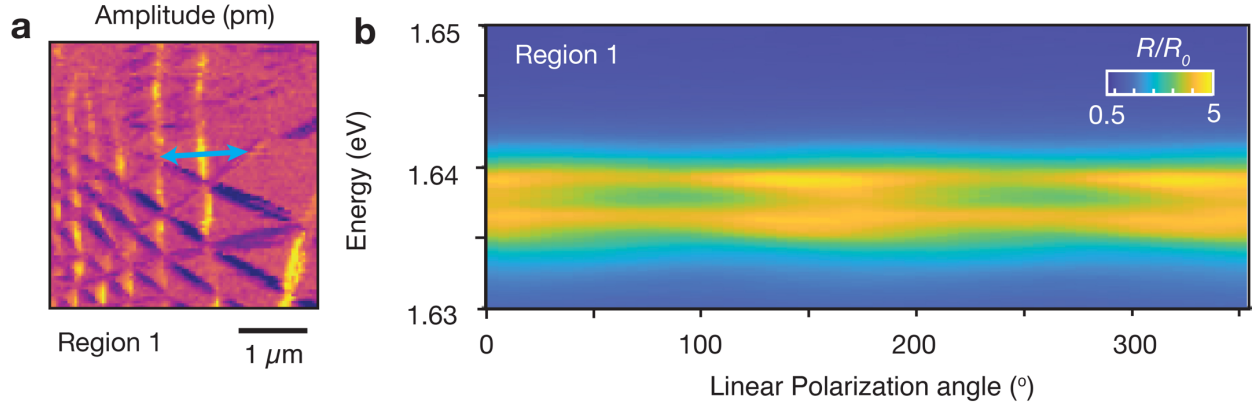

**Figure S2**. **Polarization dependence of excitons in region 1. a,** PFM amplitude map of region 1. **b,** Normalized reflectance as a function of the excitation linear polarization angle in region 1as indicated by the blue arrow in **a**. Comparing with the region 3 shown in **Fig. 3**, both the dominant domain wall and the linear polarization directions are roated by ~30 degrees in real space. In other words, the linear polarization direction remains parallel to the domain wall direction.

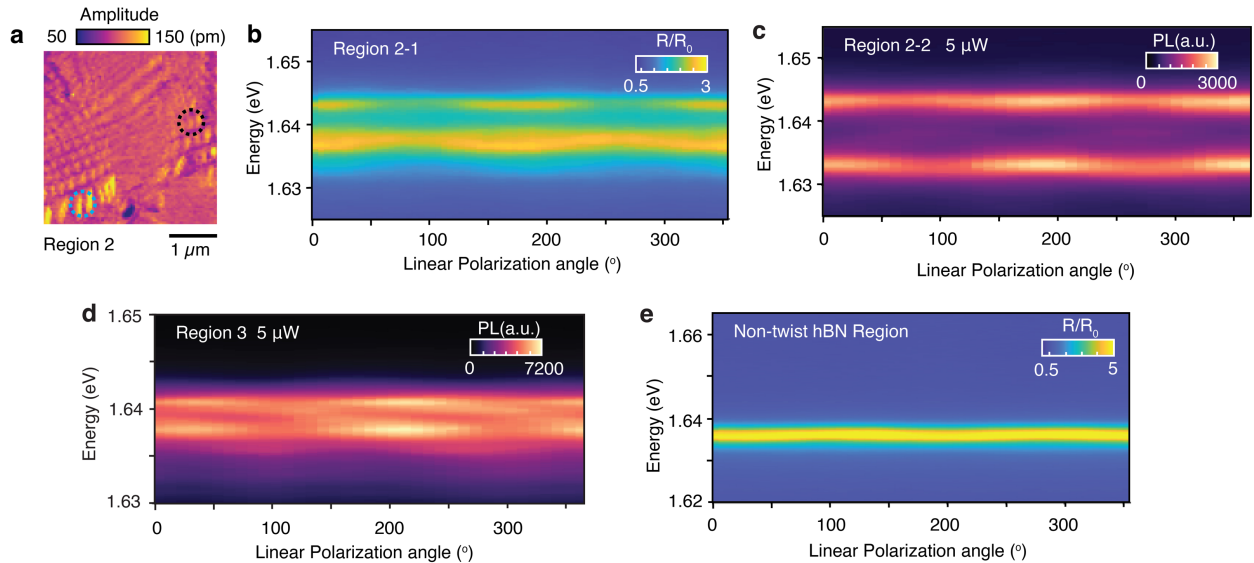

**Figure S3**. **Polarization dependence of excitons in regions with sub-diffraction domains**. **a**, PFM amplitude map of region 2. The blue and black dashed circles correspond to the detection spot in **b** and **c,** respectively. **b,** Linear polarization dependence of the exciton reflection in region 2-1. **c,** Linear polarization dependence of PL in regions 2-2, which has a small domain size of smaller than 200nm. The splitting energy of the two resonances is around 10 meV. **d,** Linear polarization PL emission for the similar spot in Fig. 3b(region 2). As the laser spot detects two domains simultaneously, there are two maxima in the PL polarization angle, which are separated

by a nearly 60°. **e**, Linear polarization dependence of the reflectance taken at the non-twist hBN region. No obvious polarization dependence is observed.

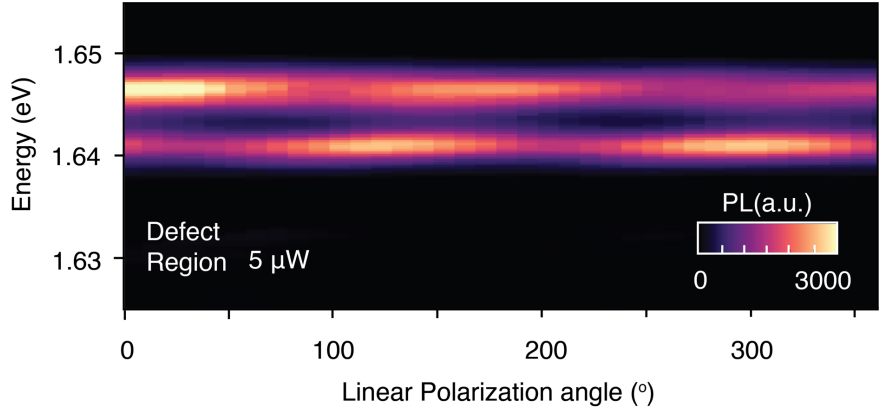

**Figure S4. Polarization dependence of energy-split excitons caused by strain.** In regions where local strain splits the exciton peaks, the two exciton peaks do not have the same linear polarization.

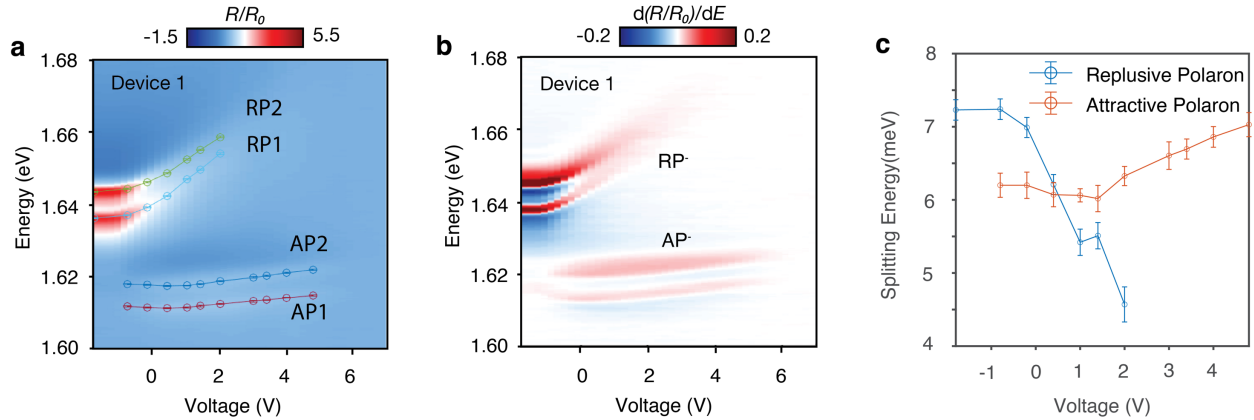

**Figure S5**. **Confinement of repulsive and attractive polarons.** Doping dependence of **a,** normalized reflectance, and **b**, its derivative with respect to energy. **c,** The energy splitting of the repulsive and attractive polarons exhibit different dependence on the doping level. The error bars represents the uncertainty in determining the peak position by Gaussian fit.

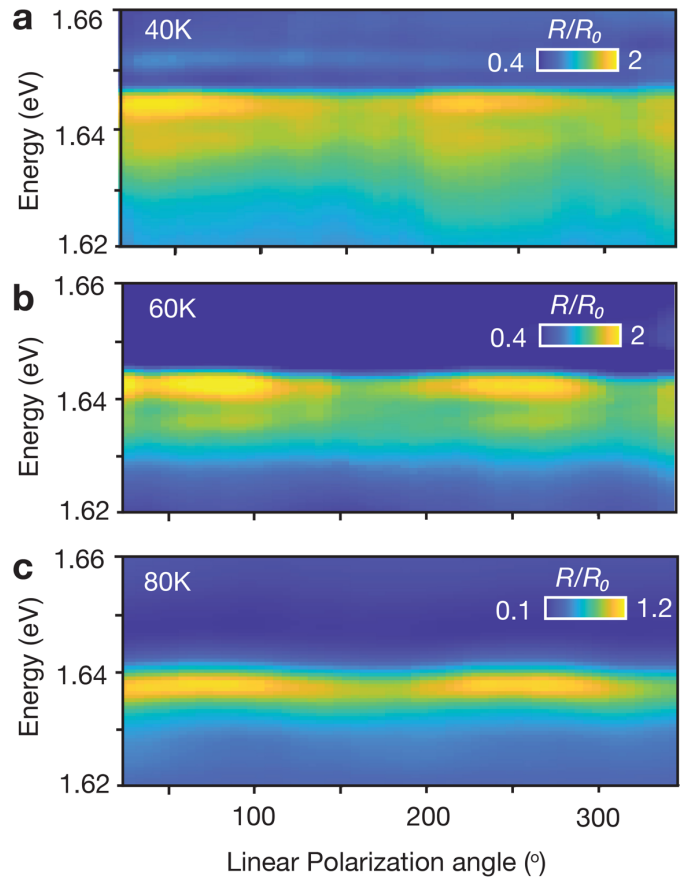

**Figure S6**. **Temperature-dependent linear polarization in D2. a-c,** Normalized reflectance as a function of the excitation linear polarization angle in D2 at 40 K, 60 K and 80 K.

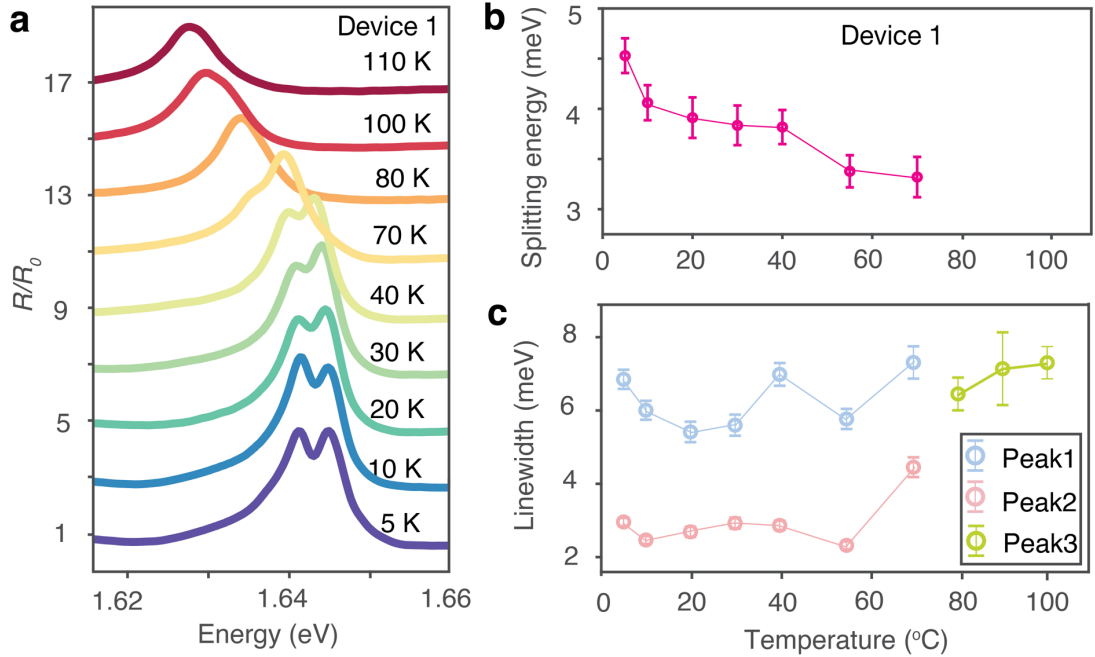

**Figure S7. Temperature-dependent quantum confinement. a**, Normalized reflectance of the intralayer excitons at different temperatures. With increasing temperature, each spectrum is shifted by a constant. **b**, The splitting energy decreases with increasing temperature and eventually vanishes near 80 K. **c**, The extracted linewidth as a function of temperature. Below 100 K, the linewidth does not change significantly, suggesting that the observed reduction in the splitting energy is not simply due to the linewidth change. The error bars represent the uncertainty associated with the fitted parameters of a Lorentzian model.